\documentclass[aps, prl, twocolumn, showpacs, superscriptaddress]{revtex4}

\usepackage{graphicx} 
\usepackage{dcolumn} 
\usepackage{bm} 
\usepackage{amsmath}
\usepackage{SIunits}
\usepackage{times}
\usepackage[ansinew]{inputenc}
\usepackage{upgreek}

\newcommand{\overbar}[1]{\mkern 1.5mu\overline{\mkern-1.5mu#1\mkern-1.5mu}\mkern 1.5mu}

\begin{document}

\title{Signatures of phase-coherent transport and the role of quantum fluctuations in the dynamical Coulomb blockade regime}

\author{Berthold J\"ack}
\email[Corresponding author; electronic address:\ ]{berthold.jack@alumni.epfl.ch}
\thanks{Present address:\ Princeton University, Joseph Henry Laboratory at the Department of Physics, Princeton, NJ 08544, USA}
\affiliation{Max-Planck-Institut f\"ur Festk\"orperforschung, 70569 Stuttgart, Germany}
\author{Jacob Senkpiel}
\affiliation{Max-Planck-Institut f\"ur Festk\"orperforschung, 70569 Stuttgart, Germany}
\author{Markus Etzkorn}
\affiliation{Max-Planck-Institut f\"ur Festk\"orperforschung, 70569 Stuttgart, Germany}
\author{Joachim Ankerhold}
\affiliation{Institut f\"{u}r Komplexe Quantensysteme and IQST, Universit\"{a}t Ulm, 89069 Ulm, Germany}
\author{Christian R. Ast}
\affiliation{Max-Planck-Institut f\"ur Festk\"orperforschung, 70569 Stuttgart, Germany}
\author{Klaus Kern}
\affiliation{Max-Planck-Institut f\"ur Festk\"orperforschung, 70569 Stuttgart, Germany}
\affiliation{Institut de Physique de la Mati{\`e}re Condens{\'e}e, Ecole Polytechnique F{\'e}d{\'e}rale de Lausanne, 1015 Lausanne, Switzerland}
\date{\today}

\begin{abstract}

Josephson junctions operated in the dynamical Coulomb blockade regime recently gained an significant amount of attention as central building block in concepts to demonstrate the non-Abelian character of Majorana fermions. Its physical properties are strongly affected by the intimate interplay of intrinsic quantum fluctuations and environmentally-induced quantum fluctuations each of which promoting different Cooper pair transport mechanisms at small voltages around zero. To shed light on the detailed transport mechanisms occurring in this type of junction, we performed voltage-biased measurements on the small-capacitance Josephson junction of a scanning tunneling microscope at milli-Kelvin temperatures. The low voltage-regime of experimental current-voltage characteristics can be modeled by the two complementary descriptions of phase coherent and incoherent Cooper pair transport, signaling the occurrence of qualitatively different transport mechanisms. This observation receives further support from analyzing the calculated Fano factor of the current noise as a probe for correlations in Cooper pair transport, following a theoretical proposal. Together our experimental observations and related data analysis provide a clear signature of coherent Cooper pair transport along with the absence of perfect charge quantized transport around zero voltage, as well as of incoherent Cooper pair transport towards higher voltages.
\end{abstract}

\maketitle

\section{Introduction}
The phase $\varphi$ of a superconducting tunnel junction connects two superconducting order parameters, $\Delta_{\rm1}$ and $\Delta_{\rm2}$, respectively, across an electrically insulating barrier. It represents a macroscopic degree of freedom and forms the basis for a plethora of phenomena, most prominently the dc-Josephson effect \cite{Anderson_1964}. Technological applications include Superconducting Quantum Interference Devices (SQUIDs) \cite{Clarke_2006} as well as Cooper pair boxes and phase qubits as building blocks for quantum information processing with superconducting platforms \cite{Devoret_2013}.

Due to its macroscopic nature, in an actual circuit the phase is inevitably subject to environmental degrees of freedom that can often be considered as thermal reservoirs \cite{Ingold_1992}. As a consequence, quantum mechanically, the properties of a superconducting junction are not only determined by the fluctuations of the phase and its conjugate partner, the charge operator $Q$, but also by the fluctuations induced by these reservoirs. While the former ones, $\delta\phi$ and $\delta Q$, depend on the respective energy scales, namely, the coupling energy $E_{\rm J}$ and the charging energy $E_{\rm C}=(2e)^2/2C_{\rm J}$ with the intrinsic capacitance $C_{\rm J}$ and elementary charge $e$, the latter are typically characterized by an environmental impedance $Z(\nu)$ through the fluctuation-dissipation theorem.

\begin{figure}
\centering
\includegraphics{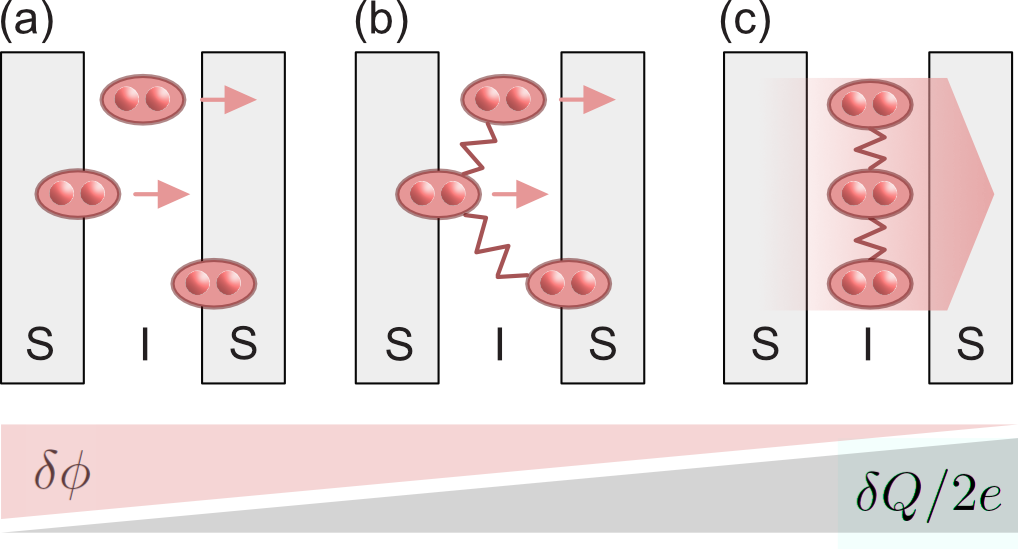}
\caption{Cooper pair transfer mechanisms occurring between two superconducting electrodes, labelled {\em S} that are separated by an insulating barrier, labelled {\em I}. (a) When quantum fluctuations of the phase are large, $\delta\phi\gg\delta Q$ at $E_{\rm C}/E_{\rm J}\gg1$, phenomena associated with a fully quantized charge, such as the dynamical Coulomb blockade, can be observed. (b) Intermediate regime where quantum phase fluctuations $\delta\phi$ and charge fluctuations $\delta Q$ exist. This regime is often realized in a low impedance environment $\rho\ll1$ with strong system-reservoir coupling. The junction properties are dominated by quantum phase diffusion and incoherent Cooper pair tunneling, respectively as demonstrated in the course of this manuscript. (c) In the absence of system-reservoir coupling ($\rho\geq1$) at $E_{\rm J}/E_{\rm C}\gg1$, quantum-phase fluctuations of the phase $\delta\phi$ are largely suppressed, $\delta\phi\ll\delta Q$, and transport is dominated by coherent Cooper pair tunneling. Occasionally, phase slips due to macroscopic quantum tunneling occur though, influencing the transport characteristics of the tunnel junction \cite{Devoret_1985}.}
\label{fig_1}
\end{figure}

In the simplest case of an ohmic impedance $R_{\rm DC}=Z(0)$ at temperature $T$, two limiting cases are well-understood for $\rho=R_{\rm DC}/R_{\rm Q}\gg1$ (Resistance quantum $R_{\rm Q}=h/(2e)^2$ and Planck's constant $h$): If $E_{\rm C}$ is the leading energy scale, i.e. $E_{\rm C}\geq E_{\rm J}\gg k_{\rm B} T$ where $k_{\rm B}$ denotes Boltzman's constant, inelastic tunneling of individual Cooper pairs occurs with a strong suppression of charge transfer in the low voltage regime, a phenomenon known as dynamical Coulomb blockade (DCB) \cite{Averin_1990, Devoret_1990}. This in turn implies that individual charging events are promoted,  suppressing charge fluctuations $\delta Q$, while the phase is strongly delocalized $\delta\phi\gg\delta Q$; see Fig.\,\ref{fig_1}(a). In the opposite domain, where $E_{\rm J}$ dominates, the phase is an almost good quantum number, i.e. $\delta\phi\ll\delta Q$, and coherent (correlated) Cooper pair transport across the tunnel barrier can be observed (Fig.\,\ref{fig_1}(c)). However, occasionally phase slips due to Macroscopic Quantum Tunneling (MQT) appear for low bias altering the tunnel junction characteristics \cite{Devoret_1985}.

Most of the actual circuits contain a low impedance environment, $\rho\ll1$ also referred to as the overdamped regime, where the coupling to the environment is strong; see Fig.\,\ref{fig_1}(b). Here, signatures of DCB are commonly observed in voltage-biased experiments with microscopic Josephson junctions (JJ) at $E_{\rm C}\geq E_{\rm J}\gg k_{\rm B} T$, where this transport mechanism manifests, for instance, as spectral resonances in the current-voltage characteristics (IVC) \cite{Grabert_1994, Holst_1994, Hofheinz_2011, Jaeck_2015}. However, small-capacitance JJs have been predicted to also support signatures of phase coherent transport at very small voltages \cite{Ingold_1999}. Quantum fluctuations of the charge $\delta Q$ are thus strongly energy dependent and basically quenched only outside the low energy sector. Indeed, a complementary picture has been revealed for quantum phase fluctuations induced from the environment $Z$ at $\rho\ll1$: Pronounced phase fluctuations emerge with increasing ratio $E_{\rm C}/ k_{\rm B} T$ and the phase dynamics turns into a quantum phase diffusion, a process which describes the transition towards the domain of DCB \cite{Ankerhold_2004, Ankerhold_2007, Jaeck_2017}. 
 
Accordingly, the changeover from a coherent to an incoherent transport regime displays the intimate interplay of intrinsic quantum fluctuations and environmentally induced fluctuations in devices consisting of JJs. Notably, a clear experimental observation has been elusive so far, mainly due to the challenge to access the right experimental parameters and to realize the right detection scheme. Nevertheless, the subject has gained renewed interest in the context of circuit architectures to create and to sense non-Abelian Majorana  quasi-particles \cite{Aasen_2016, Albrecht_2016, Deng_2016}. There, the role of electromagnetic environments has not been addressed so far but, as we have discussed, may play a decisive role.

In this article, we intend to experimentally elucidate the origin of the supercurrent peak and shed light onto the transition from coherent to incoherent Cooper pair transport at small voltages. To this end, we perform voltage-biased measurements on the tunnel junction of a scanning tunneling microscope (STM) operated at milli-Kelvin temperatures \cite{Assig_2013}. The atomic scale tunnel junction of the STM embedded in a low impedance environment, $\rho\ll1$ has proven to be excellent means to address physical phenomena such as quantum and classical Brownian motion as well as sequential Cooper pair transport mechanisms \cite{Naaman_2001, Roditchev_2006, Jaeck_2017, Jaeck_2016, Ast_2016}. Experimental current-voltage characteristics are analyzed in terms of coherent and incoherent transport. To quantify the relevance of quantum phase fluctuations, we study the energy-dependence of the Fano factor for the current noise, which we calculate from an analytic expression for the current-voltage characteristics as proposed by theory \cite{Grabert_2002}.

\section{Theory}

\begin{figure}
\centering
\includegraphics{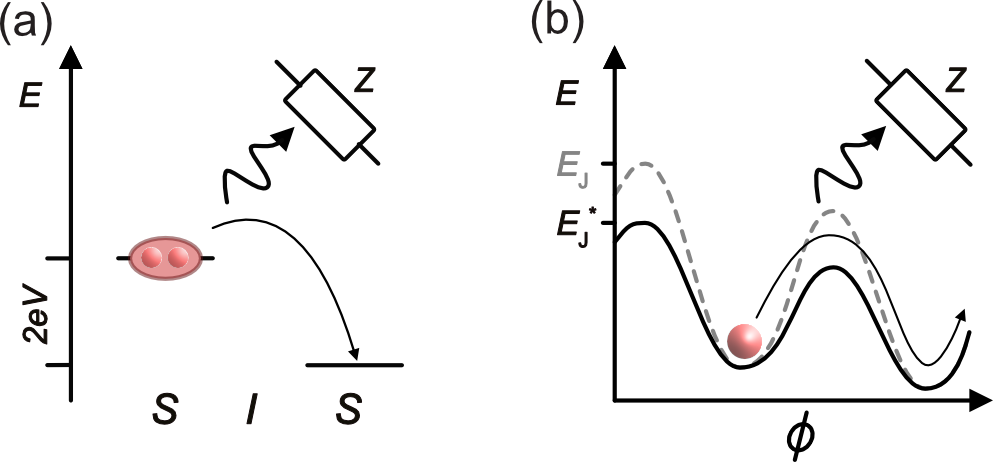}
\caption{Overdamped charge and phase dynamics in a superconducting tunnel junction. (a) Quantum phase fluctuations $\delta\phi$ induced from the electromagnetic environment $Z$ mediate the inelastic Cooper pair transfer at energy $E=2eV$. (b) In the quantum phase diffusion picture, these quantum phase fluctuations promote quasi-classical dynamics, for which $\delta\phi$ effectively reduce the washboard potential barrier height to $E_{\rm J}^*$ as compared to purely-classical dynamics with barrier height $E_{\rm J}$.}
\label{fig_2}
\end{figure}

In the dynamical Coulomb blockade regime where $E_{\rm C}\gg E_{\rm J}, k_{\rm B}T$ and for a general frequency dependent impedance $Z(\nu)$, the impact of a dissipative environment on the JJ can be analyzed by studying the Cooper pair tunneling rate across the tunnel barrier \cite{Ingold_1992, Grabert_1994}. Here, quantum phase fluctuations $\delta\phi$ induced by the dissipative environment mediate the inelastic Cooper pair transfer from one electrode to another, also see Fig.\,\ref{fig_2}(a). This process yields a measurable current $I_{\rm J}$ at finite voltages $V$ \cite{Ingold_1992, Grabert_1994}, which can be described by
\begin{equation}
\label{POET}
I_{\rm J}(V)=\frac{\pi e E_{\rm J}^{2}}{\hbar}[P(2eV)-P(-2eV)].
\end{equation} 
The $P(E)$ function reflects the probability for a transfer of energy $\pm E\propto\delta\phi$ between junction and environment. Hence, it corresponds to the probability of a Cooper pair to tunnel at the voltage $V=E/2e$. The $P(E)$ function can be obtained for an arbitrary environment $Z(\nu)$ by solving a self-consistent integral equation and, thus current-voltage characteristics can be calculated \cite{Ingold_1991, Jaeck_2016}. 

For not too large tunnel coupling values $E_{\rm J}$ and tunneling rates $P(E)$ such that $E_{\rm J}P(E)\ll1$, Eq.\,\ref{POET} can be applied to describe the Cooper pair tunneling current at voltages larger than a critical voltage $V_{\rm C}=(2e/\hbar)R_{\rm DC}E_{\rm J}^{2}/E_{\rm C}$. In experiments on tunnel junctions with capacitances of a few femto farads, $V_{\rm C}$ typically takes estimated values well below 1\,$\upmu$eV. Below this voltage, the perturbative $P(E)$-description, which only includes terms up to second order, diverges and higher order terms, which take co-tunneling events into account, may need to be included; cf. Ref. \cite{Ingold_1999}.
 
Alternatively, one can account for the effects of the dissipative environment $Z(\nu)$ on the junction characteristics by studying the quantum dynamics of the phase $\phi$ in the framework of the quantum-Smoluchowski (QSM) equation \cite{Ankerhold_2001}. Here, strong damping at $\omega_{\rm J}$ is realized by a low impedance environment, $\rho\ll1$ imposing quasi-classical phase dynamics, however, with significant correction due to quantum phase fluctuations at $E_{\rm C}\gg E_{\rm J}>k_{\rm B}T$. These quasi-classical dynamics correspond to the diffusive motion of the phase through the washboard potential landscape, as shown in Fig.\,\ref{fig_2}(b), where quantum phase fluctuations $\delta\phi$ effectively reduce the washboard potential barrier height \cite{Ankerhold_2004}. Observing QSM dynamics of the phase demands the dimensionless friction $\eta$ to sufficiently exceed the dimensionless inverse temperature $\Theta$ as well as the latter to be much larger than unity \cite{Ankerhold_2007}:

\begin{equation}
\label{QSMreq}
\eta\equiv\frac{E_{\rm C}}{2 \pi^2 \rho^2 E_{\rm J}} \gg \Theta\equiv\frac{\beta E_{\rm C}}{2\pi^2\rho}\gg1. 
\end{equation}

For an ohmic environment, $R_{\rm DC}=Z(0)$, it is possible to solve the equation of motion of the quantum-mechanical phase and derive a compact analytic expression for the Cooper pair current \cite{Grabert_1998, Ankerhold_2004}, reading
\begin{equation}\label{IVC}
I_{\rm J}(V)= \frac{e\rho\beta\pi}{\hbar}\,  (E_{\rm J}^*)^2\, \frac{\beta e V}{(\beta e V)^2+\pi^2 \rho^2}.
\end{equation} 
Quantum phase fluctuations reduce the tunnel coupling $E_{\rm J}$ to an effective value of
\begin{equation}\label{IZ2}
E_{\rm J}^*= E_{\rm J}\, \rho^{\rho}\left(\frac{\beta E_{\rm C}}{2\pi^2}\right)^{-\rho} {\rm e}^{-\rho c_{\rm 0}}.
\end{equation}
Here, $c_{\rm 0}=0.5772\ldots$ denotes Euler's constant. In contrast to the classical Josephson effect occurring strictly at zero voltage, the quantum phase fluctuation yield a non-zero phase velocity corresponding to a non-zero voltage drop across the tunnel junction, $\langle\dot{\phi}\rangle\propto V\neq0$. 

We note that for purely classical dynamics, i.\,e. at large enough temperatures where environmentally-induced quantum fluctuations can be neglected, Eq.\,\ref{IVC} corresponds to the well-known Ivanchenko-Zil'berman model where diffusion is dominated by thermally-activated phase diffusion through the potential landscape \cite{Zilberman_1969}. In this situation the renormalized tunnel coupling $E_{\rm J}^*$ is simply replaced by its bare value $E_{\rm J}$ and, in the washboard potential analogue, the original barrier height $E_{\rm J}$ is restored; cf. Fig.\,\ref{fig_2}(b). On the other hand, if environmentally-induced quantum fluctuations were absent $\delta\phi=0$, i.\,e. the system-reservoir coupling is weak at $\rho\geq1$, diffusion in the regime $E_{\rm C}\geq E_{\rm J}\gg k_{\rm B}T$ would be dominated by MQT instead, and the quantum Smoluchowski description is not applicable anymore \cite{Devoret_1985}.

For a pure ohmic environment, the Cooper pair current described by Eqs.\,\ref{POET} and \ref{IVC}, also referred to as supercurrent peak, is the hallmark feature of small-capacitance superconducting tunnel junctions. It can be typically observed at small finite voltages in voltage-biased measurements at ultra-low temperatures \cite{Steinbach_2001, Jaeck_2016}. While the supercurrent peak has been successfully described by phase and charge dynamics in different studies and different regimes \cite{Steinbach_2001, Jaeck_2016, Jaeck_2017}, the underlying physical phenomena remain elusive so far. In particular for the regime of $\rho\ll1$ and $E_{\rm C}\geq E_{\rm J}\gg k_{\rm B}T$, theory suggests the supercurrent peak to originate from the qualitatively different transport mechanisms of overdamped quantum phase diffusion and incoherent Cooper pair tunneling at different voltages \cite{Ingold_1999}. To answer this intriguing question, it was proposed to employ the Fano factor of Poissonian shot noise in order to identify the transport mechanism leading to the supercurrent peak \cite{Grabert_2002}, as we will explain in the following. 

The current shot noise represents a valuable quantity that has been successfully employed to study transport in nano- and mesoscopic systems \cite{Buettiker_2000}, such as fractionalized charge in fractional quantum Hall states \cite{Saminadayar_1997} or spin-dependent transport through single atoms in STM experiments \cite{Burtzlaff_2015}. 
For shot noise $S_{\rm I}$, it can be analyzed in terms of the Fano factor reading \cite{Buettiker_2000}, 
\begin{equation}
\label{Fano}
F_{\rm I}=S_{\rm I}/4eI.
\end{equation}
While the Fano factor is mostly accessed by measuring the current noise spectrum, it can also be obtained analytically from the current-voltage characteristics, as has been theoretically shown for small-capacitance tunnel junctions operated in the DCB regime \cite{Grabert_2002}. Here, the shot noise of the Cooper pair current $S_{\rm I}$ can be generally expressed by \cite{Grabert_2002}
\begin{equation}
\label{FF}
S_{\rm I}(V)=\frac{2e}{1-\rho}\left(I - V \frac{\partial I}{\partial V}\right).
\end{equation}
Given an analytic expression of the current-voltage characteristics $I(V)$ as well as knowledge on the normalized environmental DC impedance $\rho$, the Fano factor of the Cooper pair current noise can be directly calculated. This concept may not be generalized to other experiments. Yet, in the present case it allows us to access the Fano factor in a much easier way as compared to its direct measurement by using low temperature noise spectroscopy.

Following this theoretical proposal \cite{Grabert_2002}, we will employ the direct relation between $F_{\rm I}$ and $I(V)$ as a probe to identify the underlying transport mechanism of the supercurrent peak: In the limit $F_{\rm I}\rightarrow0$, charge transport is continuous, displaying phase-coherent transport, while for $F_{\rm I}\rightarrow1$, the noise approaches the Poissonian limit of shot noise, signaling sequential charge transport that is incoherent Cooper pair tunneling; cf. Fig\,\ref{fig_1}. In this sense, the Fano factor can be considered as a good measure for quantum phase fluctuations, which are the driving force of the transition from coherent to incoherent transport in small capacitance Josephson junctions \cite{Ingold_1992}.

\section{Experiment}
We employ the atomic scale tunnel junction of a STM operated at a base temperature of 15\,mK to study the transport characteristics of a Josephson junction in the DCB regime \cite{Assig_2013}. The tunnel junction is formed by the ultra-high vacuum (UHV) gap between a superconducting vanadium tip and a superconducting vanadium (001) single crystal \cite{Jaeck_2016}. The tip is cut in air from 250$\upmu$m thin amorphous vanadium wire of purity $3N$ and immediately transferred into the UHV chamber. A well developed superconducting gap at the tip apex is obtained by performing {\em in situ} field emission on the clean vanadium surface. The vanadium crystal surface is thoroughly cleaned by repeated cycles of sputtering (Ar$^+$ ions at 1\,keV) and thermal annealing to temperatures of $T\leq1200$\,K until it shows an atomically clean oxygen-assisted (5$\times$1) reconstruction \cite{Jaeck_2016}. A typical current-voltage characteristics of (IVC) the tunnel junction is displayed in Fig.\,\ref{fig_1}(a), showing a well-developed superconducting gap of tip and sample. We note that he STM tip gap $\Delta_{\rm 1}$ is reduced by approx. 57\,$\%$ compared to the sample gap $\Delta_{\rm 2}=800\pm11\,\upmu$eV, which is common occurrence in vanadium STM tips \cite{Jaeck_2015, Eltschka_2014, Eltschka_2015}.

The measurement circuit of our experiment is displayed in Fig.\,\ref{fig_3}(b). It contains a bias voltage source $V_{\rm B}$ (see below), the tunnel junction characterized by the tunnel element $I_{\rm 0}$ and the junction capacitance $C_{\rm J}$, and the environmental impedance $Z(\nu)$. We separate the time-scales of the biasing circuit, $\tau_{\rm RC}\approx R_{\rm 0}C_{\rm 0}$ and junction phase, $\tau_{\rm\phi}=2\pi/\omega_{\rm 0}\approx10^{-11}\,$s by using an additional large shunt capacitor $C_{\rm 0}= 3\,$nF and load-line resistor of $R_{\rm 0}= 3.5\,$k$\Omega$ and obtain $\tau_{\rm RC}\approx10^{-5}\gg\tau_{\rm\phi}$  \cite{Jaeck_2015, Joyez_1999}. In this way, we decouple the junction phase dynamics occurring at $\omega_{\rm 0}$ from the bias circuit elements such that the damping of $\phi$ is solely determined by the environmental impedance $Z(\nu)$. Concerning $Z(\nu)$, its DC part is dominated by the coupling of the tunneling Cooper pairs to the electromagnetic vacuum in the gap between tip and sample (tunnel barrier); the STM being operated in UHV. Thus, the vacuum impedance $R_{\rm DC}=Z(\nu\to 0)=377\,\Omega$ determines the DC impedance in the vicinity of the tunnel junction whereas the resistance of the leads (transmission lines) is negligibly small \cite{Jaeck_2016}. Additionally, we obtain an effective DC impedance for low GHz frequencies by choosing an STM tip of adequate length. This moves the tip resonance modes in $Z(\nu)$ to $\nu>$10\,GHz, as is illustrated in Fig.\,\ref{fig_3}(c) \cite{Jaeck_2015} and we can apply Eq.\,\ref{IVC} to describe IVCs small voltages around zero in our experiment \cite{Grabert_1994}.

\begin{figure}
\centering
\includegraphics{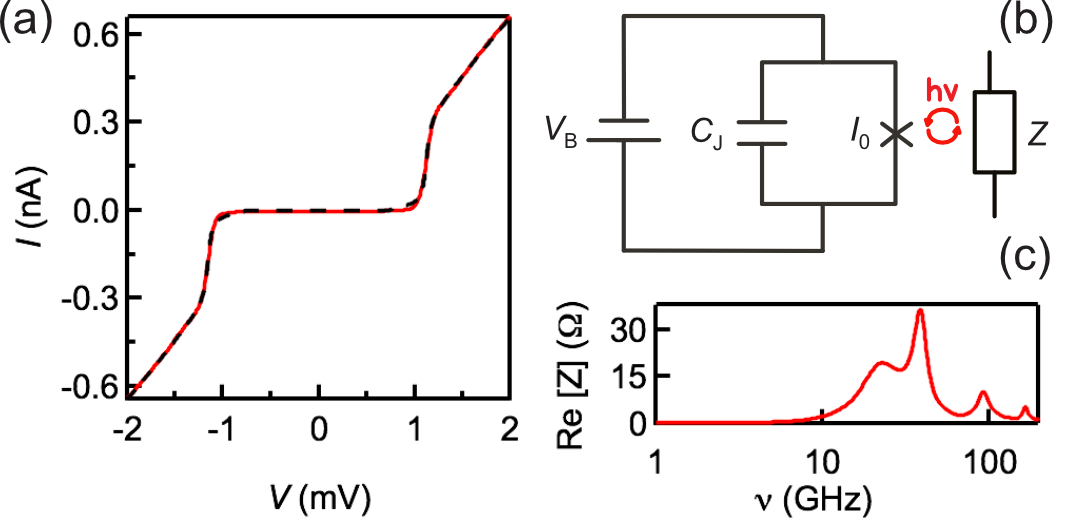}
\caption{Experimental setup. (a) Experimental IVC (red, solid) measured at $R_{\rm T}=2.80$\,M$\Omega$ and extended Dynes fit (black, dashed) \cite{Ast_2016}. (b) Simplified circuit diagram containing the voltage source $V_{\rm B}$, the environmental impedance $Z$, the junction capacitance $C_{\rm J}$ and the tunnel coupling element $I_{\rm 0}$. (c) Frequency-dependent part of $Z$ as obtained from simulations \cite{Jaeck_2015}.}
\label{fig_3}
\end{figure}

All transport measurement have been carried out by using a source measure unit with accuracies of 0.015\,$\%$ for voltage values and equal or below 0.1\,$\%$ for current values in the ranges used, with resolution limits of 100\,nV and 10\,fA, respectively. We can tune the normal state resistance $R_{\rm T}$ of the tunnel junction by changing the vacuum gap width between STM tip and sample. We note that this procedure leaves the environmental impedance $R_{\rm DC}$ unchanged, whose DC properties are dominated by the impedance of the surrounding vacuum. Before analyzing the data, the experimental IVCs are corrected for voltage drops across the low-impedance load lines as well as an experimental voltage offset from zero. After this correction has been performed, we determine the $R_{\rm T}$ values with a relative standard deviation of approx. 5\,$\%$ by fitting a linear function with slope $dI/dV=1/R_{\rm T}$ to the normal conducting part of the tunnel spectrum far outside the superconducting gap.

\section{Results}

\begin{figure*}
\centering
\includegraphics{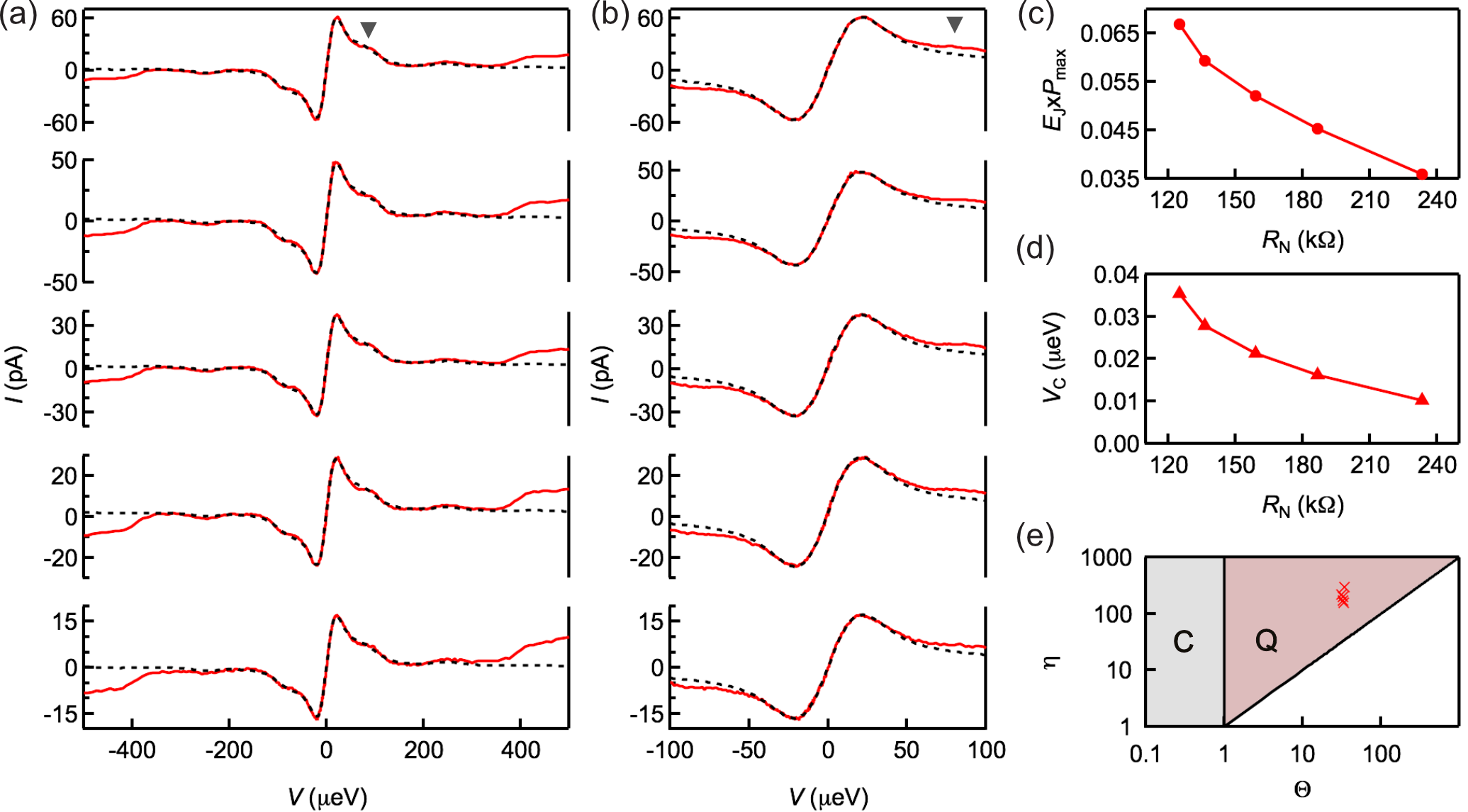}
\caption{Current-voltage characteristics in the dynamical Coulomb blockade regime. Experimental data (red, solid) and fit to the data (black, dashed) using the $P(E)$ model of Eq.\,\ref{POET} in (a) as well as the QSM model of Eq.\,\ref{IVC} in (b) for different normal state resistance values $R_{\rm T}=\{125, 136, 159, 187, 233\}\,$k$\Omega$ from top to bottom. The triangle indicates the lowest voltage, at which an environmentally-induced current resonance is experimentally observed. (c) Weak coupling condition for $P(E)$ theory $E_{\rm J}\times P(E)_{\rm max}$ as a function of $R_{\rm T}$. (d) Critical voltage $V_{\rm C}$ calculated form the $P(E)$ fit parameters as a function of $R_{\rm T}$. (e) Phase diagram for overdamped phase diffusion with the regime of quantum phase diffusion at $\eta\gg\Theta$ and $\Theta\gg1$, labelled '{\em Q}' and the regime of classical phase diffusion at $\Theta\ll1$, labelled {\em C}. The experimentally determined $(\eta,\Theta)$-pairs are plotted as crosses in red color.}
\label{fig_4}
\end{figure*}

We have studied the transport properties of the tunnel junction by performing voltage-biased experiments at different tunnel resistance values ranging from $125$\,k$\Omega<R_{\rm T}<233$\,k$\Omega$. In Fig.\,\ref{fig_4}(a), we plot the resulting IVCs and focus on the low voltage regime inside the superconducting gap; cf. Fig.\,\ref{fig_3}(a). The IVCs feature a pronounced supercurrent peak centered at around $\pm20\,\mu$eV, as it has been reported for voltage-biased measurements in the DCB regime \cite{Jaeck_2016}. At voltages larger than 50\,$\upmu$V, the spectra exhibit additional current resonances, which result from incoherent Cooper pair tunneling via interaction with the resonance modes on the STM tip \cite{Holst_1994, Hofheinz_2011, Jaeck_2015}. While those resonances represent an unambiguous signature of incoherent Cooper pair transport \cite{Grabert_1994}, we recall that the precise origin of the supercurrent peak in the DCB regime, here centered at $V\approx21\,\upmu$eV, is at question. To shed light on the involved transport mechanisms, we will, in the next step, analyze the supercurrent peak by applying the incoherent and coherent Cooper pair tunneling models of Eq.\,\ref{POET} and \ref{IVC}, respectively. 

We first analyze the IVC shown in Fig.\,\ref{fig_4}(a), by fitting their voltage dependence with $P(E)$ theory from Eq.\,\ref{POET}. To this end, we approximate the simulated impedance with a modified transmission line impedance model; see Refs.\,\cite{Grabert_1994, Jaeck_2016}. As can be seen, the fits precisely model both the supercurrent peak and the current resonances \cite{Jaeck_2016}. The fitting parameters are summarized in Table\,\ref{fit_table}. We find all values of $E_{\rm J}$, $E_{\rm C}$ and $T$ to be consistent with reported values in comparable experiments \cite{Assig_2013, Jaeck_2016, Ast_2016, Randeria_2016}. 

We find the weak coupling condition for $P(E)$ theory fulfilled for all measurements as shown in Fig.\,\ref{fig_4}(c), $E_{\rm J}P_{\rm max}<0.07$ \cite{Ingold_1992}. $P_{\rm max}=10342\,{\rm eV}^{-1}$ denotes the maximum of the $P(E)$ distribution close to zero bias (not shown) \cite{Ingold_1992, Jaeck_2016}. Additionally, we calculate the critical voltage $V_{\rm C}$ below which $P(E)$ theory is not applicable anymore \cite{Ingold_1999}. Using the fitted parameters and the relation for the critical Josephson current $I_{\rm 0}=(2e/\hbar)E_{\rm J}$, we find that except for a very small voltage range $V_{\rm C}<0.1\upmu$\,V $P(E)$ theory can be used to describe the entire IVCs in the DCB regime; see Fig.\,\ref{fig_4}(d). We note that this finding independently validates previous reports on Josephson scanning tunneling microscopy, which employ $P(E)$theory evaluated in the limit $\omega\rightarrow0$ to fully describe Cooper pair tunneling spectra \cite{Ingold_1991, Jaeck_2016, Randeria_2016}.

\begin{table}
\caption{Parameters from fitting the experimental IVCs using $P(E)$-theory and $qIZ$-theory, as shown in Fig.\,\ref{fig_3}(a) and (b), respectively.}
 	 \label{fit_table}
  \begin{tabular}{ l | c | c | c }
     & $P(E)$-theory & $qSM$-theory & $R_{\rm T}$ (k$\Omega$)\\ \hline
     & 6.47$\pm$0.01 & 6.78$\pm$0.02 & 125\\
     & 5.73$\pm$0.01 & 5.97$\pm$0.01 & 136\\
    $E_{\rm J}$ ($\upmu$eV) & 5.03$\pm$0.01 & 5.24$\pm$0.01 & 159\\
     & 4.38$\pm$0.01 & 4.49$\pm$0.01 & 187\\ 
     & 3.48$\pm$0.01 & 3.67$\pm$0.01 & 233\\ \hline
    $\overbar{E_{\rm C}}$ ($\upmu$eV) & 217$\pm$15 & 70$\pm$2 &  \\ 
    $\overbar{T}$ (mK)& 21.6$\pm$0.5 & 21.6 & \\ \hline
  \end{tabular}
\end{table}

Complementary, we employ the quantum phase diffusion description of Eq.\,\ref{IVC} to fit the supercurrent peak of the experimental data in Fig.\,\ref{fig_4}(b). As can be seen, all parts of the experimental supercurrent peak are precisely reproduced by the fit up to voltages of approximately 50\,$\upmu$V, where the first current resonance appears. We recall that Eq.\,\ref{IVC} has been derived for a pure ohmic environment. We note that for fitting the data, we convolved the calculated {\em IVC} curve from Eq.\,\ref{IVC} with a normalized Gaussian function $P_{\text{N}}(E)$ of width $\sigma=\sqrt{2E_{\rm C}k_{\rm B}T}$ to account for thermal voltage fluctuations on the junction capacitance \cite{Jaeck_2016}. Additionally, due to the mutual interplay between temperature and capacitance value in our model, we kept the temperature constant at the previously determined value from $P(E)$ theory. We employ the parameters of the QSM fits, which are summarized in Table\,\ref{fit_table}, to confirm that the junction is operated in the QSM regime. In Fig.\,\ref{fig_4}(e), we plot the calculated $(\eta,\Theta)$-pairs into the phase diagram for overdamped phase dynamics and find that the condition for QSM dynamics, $\eta\gg\Theta\gg1$ is fulfilled with $\eta>150$ and $\Theta>30$ for all $R_{\rm T}$ values. 

In the next step we compare the parameters obtained from QSM analysis with the parameters from the $P(E)$ analysis both shown in Table\,\ref{fit_table}. We find very good quantitative agreement in terms of the coupling energy $E_{\rm J}$, with the values obtained from QSM fits being slightly larger. By contrast, we observe larger mutual deviations in the fitted Coulomb energy from both models, even though both values are in the range of reported values for our experiment \cite{Jaeck_2016, Ast_2016, Jaeck_2017}. These deviations in $E_{\rm C}$ may be linked to deviations from a perfect ohmic behavior of our environmental impedance $Z(\nu)$ in the low voltage limit $V\rightarrow0$, as is required by the QSM model, cf. Fig.\,\ref{fig_2}(c). Another error source may be found in the $P(E)$ model. Deviations in the simulated impedance $Z(\nu)$, cf. Fig.\,\ref{fig_2}(c), as compared to the real impedance occurring in our experiment, results in a less accurate $P(E)$ fit. This is illustrated in Fig.\,\ref{fig_3}(a), where residual deviations between $P(E)$ fit and experimental data can be observed for the current resonances, eventually yielding less accurate fitting parameters.

\section{Discussion}

\begin{figure}
\centering
\includegraphics[width=0.5\textwidth]{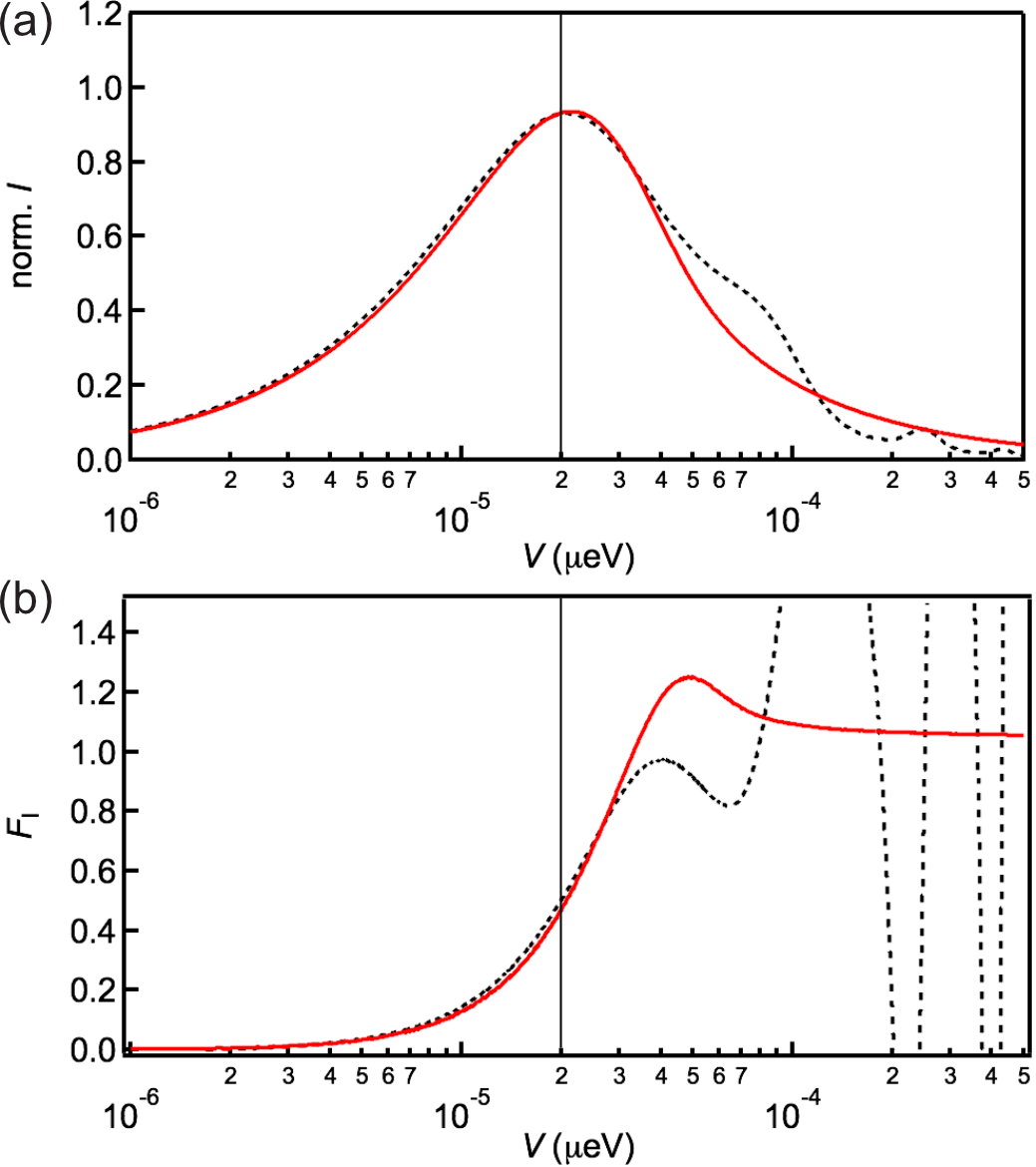}
\caption{Identification of transport mechanism via Fano factor analysis. (a) Normalized fitted IVCs from the $P(E)$-theory (black, dashed) and QSM theory (red, solid) as a function of the voltage $V$, (b) Fano factor of the current noise $F_{\rm I}$, calculated from the $P(E)$ fits (black, dashed) and QSM-fits (red, solid) as a function of the voltage $V$. $V_{\rm max}$ is indicated as a thin vertical line.}
\label{fig_5}
\end{figure}

We begin our discussion by reiterating the surprising observation of the previous section that we can describe the low voltage regime of our experimental IVCs at V$\leq40\,\upmu$V in the two qualitatively different frameworks of overdamped charge and phase dynamics. For more clarity, this finding is again highlighted in Fig.\,\ref{fig_5}(a) where we compare two normalized IVCs from the $P(E)$ and QSM fits on a semi-logarithmic scale and observe nearly perfect agreement for voltages smaller than voltage of the supercurrent peak maximum, V$\leq20\,\upmu$V. It follows that at these voltages the Cooper pair transport cannot be fully incoherent as is commonly associated with the $P(E)$ theory description, despite the inherent quantum phase fluctuations at $E_{\rm C}\gg E_{\rm J}$ and strong system-reservoir coupling at $\rho\ll1$. Instead, transport exhibits some correlations between the tunneling Cooper pairs; cf. Fig.\,\ref{fig_1}(b). This observation contrasts previous studies on small-capacitance superconducting tunnel junctions \cite{Steinbach_2001}: here, a classical phase-diffusion picture sufficed to describe the supercurrent peak, which can likely be explained by a large on-chip shunt capacitance promoting classical dynamics at $\Theta\ll1$ in their experiment, cf. Eq.\,\ref{QSMreq}. To shed more light on the subtle interplay between inherent quantum phase fluctuations and quantify the relevance of environmentally-induced quantum fluctuations, we now calculate the Fano factor of the current noise $F_{\rm I}$. In the following, we will employ it as a sensitive probe to distinguish the transport mechanisms occurring at V$\leq20\,\upmu$V.

We can calculate the Fano factor $F_{\rm I}$ of the current noise by employing the previously fitted IVCs shown in Fig.\,\ref{fig_4}(a) and (b) as an input for Eqs.\,\ref{Fano} and \ref{FF}. Using both curves from the $P(E)$ and QSM fit we calculate $F_{\rm I}$, which we plot as a function of the voltage drop across the tunnel junction in Fig.\,\ref{fig_5}(b). As it directly follows from the formal relation to the current in Eqs.\,\ref{Fano} and \ref{FF}, both $F_{\rm I}$ functions match almost perfectly at V$\leq20\,\upmu$V, yet exhibit deviations to larger voltages owing to the current resonances in the $P(E)$ curve. These oscillating features relate to a sign change in the current derivative in Eq.\,\ref{FF} induced by the current resonances and, thus can be regarded as an artifact in the calculation of $F_{\rm I}(V)$.The voltage dependence of both Fano factors shown in Fig.\,\ref{fig_5}(b) can be described by two main features: We find a strong decrease from $F_{\rm I}\approx0.5$ at the supercurrent peak maximum $V=V_{\rm max}\approx20\,\upmu$eV to zero value in the limit $V\to0$, and a strong increase towards higher voltages $V>V_{\rm max}$, where $F_{\rm I}\to1$. 

Regarding our physical interpretation, the value $F_{\rm I}\approx0.5$ at $V=V_{\rm max}$ signals that Cooper pair transport of the supercurrent peak is not fully sequential anymore in this voltage range, but exhibits correlations between the tunneling pairs. In this intermediate regime depicted in Fig.\,\ref{fig_1}(b), environmentally-induced quantum fluctuations are substantial, promoting the occurrence of overdamped quantum phase diffusion and sequential Cooper pair tunneling, respectively. Accordingly, the resulting IVC can be described in the respective models as demonstrated before. In the extreme limit $0\leq V\ll V_{\rm max}$, the reduction of $F_{\rm I}\to0$ indicates fully phase-coherent transport. Here, $\delta\phi$ is fully quenched and MQT should become observable if one were able to experimentally resolve this small voltage range \cite{Ingold_1999}. Formally, this regime coincides with the voltage range $V<V_{\rm C}$, in which $P(E)$ theory derived in 2$^{\rm nd}$ order perturbation theory is not applicable \cite{Ingold_1992}. Towards higher voltages $V>V_{\rm max}$, the Fano factor approaches the Poissonian limit, $F_{\rm I}\rightarrow1$ and transport is supported by sequentially tunneling Cooper pairs. In this regime, phase coherence is strongly perturbed by the system-reservoir coupling and the environmentally-induced $\delta\phi$ mediate the incoherent Cooper pair tunneling \cite{Grabert_1994}. This latter finding has additional experimental support from the observation of current resonances in the IVCs shown in Fig.\,\ref{fig_4}(a) and has been reported before as well \cite{Holst_1994, Hofheinz_2011, Jaeck_2015}.

An important consequence of our spectral Fano factor analysis relates to the ongoing research on creating and manipulating Majorana quasiparticles (MQP) in proximitized semiconducting nano-wires \cite{Albrecht_2016, Deng_2016, Aasen_2016}. Here, transport across small-capacitance Josephson junctions operated in the DCB regime is proposed to test the existence of MQP localized at the nano-wire ends \cite{Aasen_2016}. Additionally, a quantized zero-bias current across the nano-wire originating from sequential Cooper pair tunneling is predicted to serve as evidence for the non-Abelian character of MQP, if effects of the surrounding circuit environment are neglected \cite{Aasen_2016}. The results of our study, however, indicate the existence of correlated Cooper pair tunneling in the limit $V=0$, which corresponds to the absence of perfect charge quantization in the Cooper pair transport. Eventually, this effect leads to a wash out of current quantization in longitudinal transport, which may hamper this detection scheme.

\section{Conclusion}
We have studied transport mechanisms occurring in superconducting tunnel junctions operated in the DCB regime at $E_{\rm C}\geq E_{\rm J}\gg k_{\rm B} T$ and $\rho\ll1$, where the junction properties are determined by the intimate interplay of intrinsic and extrinsic quantum fluctuations. We have performed voltage-biased measurements on the atomic-scale tunnel junction in a STM at a base temperature of 15\,mK. The experimental supercurrent peak occurring around zero voltage can be modeled by the two complementary descriptions of phase coherent and sequential Cooper pair transport. Additionally, we employ the calculated Fano factor of the Poissonian shot noise $F_{\rm I}$ as a probe for correlations in the charge transport. This helps us to identify two different transport regimes in our experiment in agreement with theory \cite{Ingold_1999, Grabert_2002}: At low voltages around the supercurrent peak maximum, the transport exhibits correlations between the tunneling Cooper pairs and charge quantization is washed out. By contrast, at larger voltages the incoherent tunneling of individual Cooper pair dominates. Our results are of general validity yet they should be of particular interest for the ongoing research on Majorana quasiparticles in hybrid superconductor-nanowire systems \cite{Aasen_2016}. We anticipate that our work stimulates further experiments in which the coexistence of coherent and incoherent Cooper pair transport will be unequivocally determined by measuring the Fano factor of the current noise in well-tailored circuits.

\section{Acknowledgements}
It is our pleasure to acknowledge inspiring discussions with D. Esteve $\&$ F. Tafuri. B. J\"{a}ck acknowledges funding from the Alexander von Humboldt foundation during the completion of the manuscript. Financial support was provided by the Center for Integrated Quantum Science and Technology ($IQ^{ST}$) (JA, CA, KK) and the German Science Foundation (DFG) through Grant No AN336/11-1 (JA).

\end{document}